\documentclass[number,sort&compress,preprint]{elsarticle}
\journal{Combustion and Flame}
\usepackage{lmodern}
\usepackage{mathtools}
\usepackage{booktabs}
\usepackage{pgfplotstable}
\pgfkeys{/pgf/fpu=true}
\pgfplotsset{compat=1.11}
\pgfplotstableset{
    every head row/.style={
        before row=\toprule,
        after row=\midrule,
    },
    every last row/.style={
        after row=\bottomrule,
    },
    /pgf/number format/sci generic={
        mantissa sep={\times},
        exponent={10^{#1}}
    },
    create on use/{sigma percent norm}/.style={
        create col/expr={\thisrow{2 sigma norm}/\thisrow{Mean Normal Temperature}*100},
    },
    create on use/{sigma percent uniform}/.style={
        create col/expr={\thisrow{2 sigma uniform}/\thisrow{Mean Uniform Temperature}*100},
    },
    create on use/{sigma percent tri}/.style={
        create col/expr={\thisrow{2 sigma tri}/\thisrow{Mean Triangle Temperature}*100},
    },
    create on use/{ind percent norm}/.style={
        create col/expr={\thisrow{delta TC normal}/\thisrow{Ind Param TC}*100},
    },
    create on use/{ind percent uni}/.style={
        create col/expr={\thisrow{delta TC uniform}/\thisrow{Ind Param TC}*100},
    },
    create on use/{ind percent tri}/.style={
        create col/expr={\thisrow{delta TC triangle}/\thisrow{Ind Param TC}*100},
    },
    create on use/{sigma dist percent}/.style={
        create col/expr={\thisrow{Monte Carlo}/\thisrow{Mean Calc TC}*100},
    },
    create on use/{delta dist percent}/.style={
        create col/expr={\thisrow{Independent Parameters}/\thisrow{Ind Param TC}*100},
    },
}
\usepackage{multirow}
\usepackage{threeparttable}
\usepackage{siunitx}
\DeclareSIUnit{\torr}{Torr}
\usepackage[version=3]{mhchem}
\usepackage{bookmark}
\hypersetup{hidelinks}
\usepackage[capitalize]{cleveref}

\DeclareMathOperator{\lam}{W}

\begin{document}
\begin{frontmatter}
\title{On the Uncertainty of Temperature Estimation in a Rapid Compression Machine}
\author{Bryan~W.~Weber\corref{cor1}}
\cortext[cor1]{Corresponding Author}
\ead{bryan.weber@uconn.edu}
\author{Chih-Jen~Sung\corref{}}
\author{Michael~Renfro\corref{}}
\address{Department of Mechanical Engineering, University of Connecticut, Storrs, CT, USA}

\begin{abstract}
Rapid compression machines (RCMs) have been widely used in the combustion literature to study the low-to-intermediate temperature ignition of many fuels.
In a typical RCM, the pressure during and after the compression stroke is measured.
However, measurement of the temperature history in the RCM reaction chamber is challenging.
Thus, the temperature is generally calculated by the isentropic relations between pressure and temperature, assuming that the adiabatic core hypothesis holds.
To estimate the uncertainty in the calculated temperature, an uncertainty propagation analysis must be carried out.
Our previous analyses assumed that the uncertainties of the parameters in the equation to calculate the temperature were normally distributed and independent, but these assumptions do not hold for typical RCM operating procedures.
In this work, a Monte Carlo method is developed to estimate the uncertainty in the calculated temperature, while taking into account the correlation between parameters and the possibility of non-normal probability distributions.
In addition, the Monte Carlo method is compared to an analysis that assumes normally distributed, independent parameters.
Both analysis methods show that the magnitude of the initial pressure and the uncertainty of the initial temperature have strong influences on the magnitude of the uncertainty.
Finally, the uncertainty estimation methods studied here provide a reference value for the uncertainty of the reference temperature in an RCM and can be generalized to other similar facilities.
\end{abstract}

\end{frontmatter}


\section{Introduction}
\label{sec:introduction}

Rapid compression machines (RCMs) are used to study the high-pressure, low-to-intermediate temperature combustion chemistry of fuels.
A comprehensive review of the design and operation of RCMs was recently given by \citet{Sung2014}.
In general, RCMs rapidly compress a fixed mass of gas to the desired pressure and temperature conditions using one or two pistons.
Subsequently, the piston (or pistons) is (are) locked in place and reactions are allowed to proceed in a constant volume chamber.
After some delay, the mixture ignites and the so-called ignition delay is one of the primary data reported from RCMs.

The ignition delays are typically reported with reference to the compressed conditions, that is, the conditions at the end of compression (EOC).
The pressure and temperature at the EOC are represented by $P_C$ and $T_C$ for pressure and temperature respectively.
The pressure is measured directly by a dynamic pressure transducer installed on the reaction chamber.
However, it is rather difficult to directly measure the temperature of the gases in the reaction chamber during and after compression.
Intrusive temperature measurement methods such as thermocouples may introduce inhomogeneities into the reaction chamber and may have time constants that are not well matched to the rate of change of temperature during compression.
In addition, non-intrusive optical techniques are difficult to set up and require extensive calibration at the pressures and temperatures of interest in RCM studies.
Thus, the temperature is determined indirectly by applying a set of assumptions collectively called the ``adiabatic core hypothesis'' to the reaction chamber \cite{Lee1998,Hu1987} (see \cref{sec:calculate-TC,sec:systematic-uncertainties}).
This approach has been previously validated by computational (e.g.\ \cite{Mittal2008b}) and experimental (e.g.\ \cite{Das2012a,Uddi2012}) approaches under some representative conditions.

\subsection{Systematic Uncertainties of RCM Experiments}
\label{sec:systematic-uncertainties}

RCMs, like all experiments, are affected by both random and systematic uncertainties.
A number of the systematic uncertainties relevant to RCMs have been identified and discussed in the literature.
Several of these are discussed in this section, while the bulk of the paper is devoted to an analysis of the random uncertainties.
The interested reader is referred to the work of \citet{Sung2014} and references therein for additional discussion.

The primary assumption affecting the uncertainty of the temperature in the RCM is the adiabatic core hypothesis.
This hypothesis states that heat loss from the reactants only occurs in a small boundary layer at the periphery of the reactor.
Thus, the majority of the gas in the reaction chamber, referred to as the ``core region'' or the ``core'', is unaffected by the heat loss and is compressed adiabatically.
Any fluid motion that would mix the boundary layer and the core gases would result in the failure of the adiabatic core hypothesis.
This includes bulk fluid motions, such as those caused by the motion of the piston, as well as random fluctuations due to turbulence.

Modern RCMs incorporate a crevice around the circumference of the piston(s) to minimize the effect of bulk fluid motions caused by the piston.
This crevice should be sized so that it can capture the boundary layer along the cylinder wall.
Without a properly sized crevice, the motion of the piston may cause mixing of the boundary layer and the core gases.
The optimal size of the crevice depends on the thickness of the boundary layer, which is in turn dependent on the specific gases used as reactants, as well as the prevailing conditions during compression \cite{Wurmel2005,Mittal2007}.

In addition, it is desired that a negligible amount of turbulence is developed during the compression stroke.
Any turbulence would likely cause mixing of the thermal boundary layer along the cylinder periphery with the core gases.
The effect of varying initial turbulence levels on ignition of syngas mixtures was investigated by \citet{Ihme2012}, who used computational solutions of a Lagrangian Fokker-Planck model to show that the compression stroke of the RCM induces a mean strain on the gases in the reaction chamber.
This mean strain had the effect of amplifying any turbulence present in the reaction chamber prior to the start of compression.
However, for smaller compression ratios, the amplification of the initial turbulence level is comparatively small \cite{Ihme2012}.

Additionally, \citet{Wurmel2005} compared simulated pressure traces for laminar and turbulent CFD calculations.
They showed good agreement between the laminar and turbulent cases for \ce{Ar}, \ce{N2}, and \ce{O2}, but relatively worse agreement for \ce{He}.
\citet{Wurmel2005} suggested that an optimized turbulence model is required for \ce{He}.
\citet{Mittal2006b} compared the results of laminar and turbulent CFD simulations with experimental results from their RCM.
They noted that, for a creviced piston, the temperature predicted by the adiabatic core hypothesis matched closely with the maximum temperature in the reaction chamber derived from CFD simulations for both laminar and turbulent cases up to approximately \SI{100}{\milli\second} after the EOC.
Nonetheless, \citet{Mittal2006b} noted that the aerodynamic features of the RCM were better predicted by laminar calculations.

\citet{Mittal2006b} also showed that the effect of the roll-up vortex could be minimized by operating at higher EOC pressures.
They showed this experimentally, by performing acetone PLIF measurements of the temperature field in the reaction chamber, as well as through CFD experiments, as described previously.
In addition, \citet{Mittal2012} conducted CFD investigations to determine the effect of the geometric parameters (i.e.\ the piston stroke length and the final clearance length), as well as the effect of the EOC pressure, on the temperature field post-compression.
They found that combinations of shorter stroke, larger clearance, and higher EOC pressure resulted in higher temperature field uniformity.

In addition, the piston velocity during compression is an important parameter in controlling the homogeneity of the mixture.
In general, one wishes to have shorter compression times such that the effects of heat loss and chemical reactions during the compression stroke are reduced.
These conditions are promoted by higher piston velocities.
However, high piston velocities may lead to increased turbulence and the breakdown of the adiabatic core hypothesis.

The choice of diluent gases can also affect the validity of the adiabatic core hypothesis.
An investigation of such effects in RCMs and shock tubes was conducted by \citet{Wurmel2007}.
They noted that using monatomic diluents results in longer ignition delays relative to diatomic diluents due to the higher thermal diffusivity of the monatomic diluents.
Moreover, \citet{Wurmel2007} showed that using helium as the diluent produces longer ignition delays of 2,3-dimethylpentane as compared to using argon as the diluent.
This effect may be attributed to the thermal conductivity of the diluent, because the molar-specific heat capacities of the monatomic diluents are equal.

\citet{Di2014} also investigated the effect of diluent composition on the ignition delay.
They considered several competing effects of the diluent gas, namely the chemical effect, the dilution effect, and the thermal effect, whereas \citet{Wurmel2007} only considered the thermal effect.
The chemical effect influences the reaction rates through the third-body collision efficiency of the diluent gas, while the dilution effect influences the ignition delay by changing the oxygen concentration; the thermal effect is due to the varying specific heat of each of the diluents.
The chemical and dilution effects should not play a large role in determining the validity of the adiabatic core hypothesis, but the thermal effect may play a role by determining the extent of the boundary layer growth during and after the compression.
Monatomic diluents may especially affect the assumption of the adiabatic core because their higher thermal diffusivity may lead to a larger boundary layer thickness, which may not be fully captured by a crevice optimized for a different diluent or different conditions.

Many RCMs are equipped with facilities to preheat the mixture and reaction chamber before it is compressed.
In general, the mixing tanks and reaction chamber are heated by tapes wrapped around the outside of each, but the piston on the inside of the reaction chamber is unheated.
The power output from the heaters is typically controlled by PID controllers with feedback from thermocouples strategically placed around the RCM and mixing tanks.
In principle, variations in the preheat temperature, either spatially or temporally, may affect the uncertainty of the temperature at the EOC.
Therefore, caution needs to be exercised when setting up the preheat control system.
Measurements of the initial temperature along the centerline of the reaction chamber for the system under analysis in this study under quiescent conditions show only a small change of temperature ($<\SI{10}{\kelvin}$) until the location of the piston.
Moreover, \citet{Wurmel2005} showed via CFD studies that inhomogeneities in the boundary temperatures (i.e.\ the reactor wall and piston temperatures) cause only a small difference in the mass-averaged temperature during and after compression compared to a case with uniform boundary temperatures.

Thus, the systematic uncertainties inherent to the design and operation of RCMs may play a large role in determining the uncertainty of the EOC temperature.
In general, these systematic effects will most likely increase the uncertainty of the temperature at the EOC compared to an estimate that does not consider them.
However, to the best of our knowledge, a detailed analysis of the uncertainty of the EOC temperature due to random errors in the experimental equipment has not been considered.
As such, the analysis in the present work considers the random errors for a suitably constructed and operated RCM under the condition that the adiabatic core hypothesis holds.
This allows the analysis to be generalized to similar experimental facilities that rely on estimations of the temperature rather than direct measurements.
Future studies should be considered to extend this analysis to include the details of systematic uncertainties.

\subsection{Methods of Uncertainty Analysis}
\label{sec:unc-methods-intro}

That the temperature must be estimated by several measured parameters and thermodynamic properties of the reactants implies that its uncertainty must be estimated by an uncertainty propagation analysis.
Such analyses have been conducted in the past, for example in the work of \citet{Weber2011} and \citet{Mittal2007a}.
Previous analyses generally assumed that the uncertainty of each parameter was normally distributed and independent (i.e.\ uncorrelated).
The parameters used for the previous analysis, e.g.\ by \citet{Weber2011}, are the same as those considered in \cref{sec:independent-parameters} in this work.
In addition, the uncertainty was estimated by linearizing the equation for $T_C$ and applying the method of quadratic sums \cite{Taylor1982}.

However, it is unclear a priori if the assumption of normally distributed, independent parameters will hold such that the quadratic sum approach can be applied.
In particular, \cref{sec:test-gas} describes the typical RCM operation procedure whereby the reactant gases are filled sequentially into the mixing tank.
Thus, the uncertainty in the partial pressure of one component cascades to affect the filled pressure of every subsequent component.
In addition, the uncertainty of each parameter is estimated from the manufacturer's specifications, but it is not clear from the manufacturer's documentation what sort of uncertainty distribution should be assumed within the specified range.
For instance, the uncertainty of many commonly used thermocouples is specified using a tolerance which must not be exceeded; however, no further information is given about how the uncertainty is distributed within this tolerance range.

Since the assumption of normally distributed, independent parameters may not hold, it would be useful to investigate alternate methods to estimate the uncertainty.
One such alternate method is a Monte Carlo analysis.
In this method, the uncertainty distribution of each of the parameters can be directly specified and the effect of different distributions on the overall uncertainty can be estimated.
Moreover, using a Monte Carlo method allows the uncertainties of parameters to be properly correlated with one another.

Thus, the objectives of this work are twofold.
The first is to describe a detailed procedure using Monte Carlo analysis to estimate the uncertainty of the compressed temperature in an RCM assuming that the adiabatic core hypothesis holds.
The Monte Carlo analysis takes into account the correlation between several of the parameters and investigates the effects of the uncertainty distributions of the parameters on the overall uncertainty in calculated temperature.
This procedure can be generalized to different RCMs and other similar facilities, provided that the assumptions made in the following are noted.
The second objective is to compare the uncertainty estimated by the Monte Carlo method to the uncertainty estimated under the assumption of normally distributed, independent parameters for a particular RCM system.
This comparison will help to determine the conditions under which it is most important to perform the detailed Monte Carlo analysis.
In addition, the assumption of normally distributed parameters will be investigated within the independent parameters analysis to determine the effect of this assumption.

\section{Typical Experimental Procedure}
\label{sec:test-gas}

The typical operating procedure of the RCM at the University of Connecticut (UConn) has been described extensively in the work of \citet{Mittal2007} and will be described briefly here for reference.
The present RCM is a pneumatically-driven/hydraulically-stopped single-piston arrangement.
The piston used to compress the reactant gases is machined with a crevice around its circumference to contain the roll-up vortex and promote homogeneous conditions in the reaction chamber.
At the start of an experimental run, with the piston in the EOC position, the reaction chamber is vacuumed to less
than \SI{1}{\torr}.
After the piston is retracted, and is locked in place by hydraulic pressure, the reaction chamber is filled with the required initial pressure ($P_0$) of test gas mixture from a mixing tank.
Finally, compression is triggered by releasing the hydraulic pressure through an electrically-operated solenoid valve.
The compression is achieved by the use of compressed nitrogen and takes approximately \SIrange{30}{50}{\milli\second}, with the majority of the pressure and temperature rise confined to the last \SI{5}{\milli\second} of the compression stroke.

Fuel/oxidizer pre-mixtures are prepared in two mixing tanks, one approximately \SI{16.6}{\liter} and the other approximately \SI{15.2}{\liter} in volume (see \cref{sec:unc-V0}).
These large volumes allow many runs to be conducted from one mixture preparation.
The mixing tanks are connected to the reaction chamber by flexible stainless steel manifold tubing.
The tanks, reaction chamber, and connecting manifold are wrapped in heating tapes and insulation to control the initial temperature of the mixture.
Temperature controllers from Omega Engineering use thermocouples---described in \cref{sec:unc-T0}---placed on the lid of each mixing tank, approximately in the center of each mixing tank, embedded in the wall of the reaction chamber, and near the inlet valve of the reaction chamber to control the preheat temperature of the mixture.
A static pressure transducer---described in \cref{sec:unc-p0}---measures the pressure in the manifold and mixing tanks.
This transducer is used during mixture preparation and to measure the initial pressure of a given experiment.

Liquid and gaseous fuels can be studied in the present RCM.
First, the mixing tanks are vacuumed to an ultimate pressure less than \SI{5}{\torr}.
If a gaseous fuel is to be studied, each of the mixture components (including the fuel(s) and any oxidizer or diluents) are added sequentially to the mixing tank at room temperature.
This procedure can also be used for liquid phase fuels whose vapor pressure at room temperature is sufficient to fill the mixing tanks to the required partial pressure.
If the liquid fuel does not provide sufficient vapor pressure at room temperature, a different procedure may be used to introduce the fuel into the mixing tanks.
In this alternate procedure, the liquid fuel is massed in a syringe prior to injection into the mixing tank through a septum.
The syringe is massed again after injection and the difference in mass from the measurement prior to the injection is taken as the injected mass.
Subsequently, proportions of \ce{O2}, \ce{N2}, and other inert gases (e.g.\ \ce{Ar}) are added manometrically at room temperature.
For liquid fuels with sufficiently high vapor pressure at room temperature, either procedure can be used; for liquid fuels with low vapor pressure at room temperature, the second procedure must be followed.
Tests have been conducted to compare these procedures for several fuels with high vapor pressure at room temperature.
In these tests, it was found that the measured partial pressure in the mixing tank closely matches with the expected partial pressure computed from the liquid mass injected into the tank.

Whichever procedure is used, the order of filling and total pressure in the tank after each gas is filled are recorded.
The preheat temperature of the RCM is set above the saturation point for the fuel to ensure complete vaporization.
The temperature inside the mixing tank is allowed to equilibrate for approximately \SI{1.5}{\hour}.
A magnetic stirrer mixes the reactants for the entire duration of the experiments.
This mixing ensures the effects of thermal and concentration gradients in the mixing tank are minimized.

Fuel blends made of several components (e.g.\ gasoline, jet fuel, or surrogate mixtures for these real fuels) can be studied in this facility by employing the same procedure and ensuring that the preheat temperature is set above the temperature required to vaporize the component with the highest saturation temperature (see e.g.\ the work of \citet{Kukkadapu2013,Kukkadapu2012a}).
In addition, other experimental procedures regarding mixture preparation have been developed; for instance, work by \citet{Allen2013} directly injected the fuel and oxidizer into the reaction chamber.
However, the focus of this study is on RCMs that use a pre-vaporization procedure in a separate mixing tank and these alternative mixture preparation methods are not considered in this work.

This approach to mixture preparation has been validated in several previous studies by withdrawing gas samples from the mixing tank and analyzing the contents by GC/MS \cite{Weber2011}, GC-FID \cite{Kumar2009}, and GC-TCD \cite{Das2012}.
The study of \citet{Weber2011} verified that the mole fraction of fuel was within \SI{5}{\percent} of the expected mole fraction and the study of \citet{Das2012} verified that the mole fraction of water was within \SI{2}{\percent} of the expected value.
In addition, both the work by \citet{Kumar2009} on \textit{n}-decane and the study of \citet{Weber2011} on \textit{n}-butanol confirmed that there was no fuel decomposition in the mixing tank over the course of a typical set of experiments.

In either of the mixing tank filling procedures described above, the uncertainties of the amount of each gaseous component are necessarily coupled because the partial pressure of each component is found by subtracting consecutive recordings of the total mixing tank pressure.
Thus, the assumption of independent parameters may not be appropriate when conducting an uncertainty analysis and the effect of the correlation should be determined for each experimental facility.

\section{Determination of Compressed Temperature}
\label{sec:calculate-TC}


According to the adiabatic core hypothesis (see \cref{sec:systematic-uncertainties}), it is assumed that the core gases in the reaction chamber are compressed isentropically.
Heat loss from the reactants during and after compression only occurs in a thin boundary layer near the wall, and the central core region is unaffected by heat loss (i.e.\ the core is adiabatic) \cite{Desgroux1995,Mittal2007,Mittal2006b}.
Thus, the heat loss is modeled as an effective reduction in the geometric compression ratio, and the temperature at the end of the compression stroke can be calculated by:
\begin{align}
\ln{\left(\frac{P_C}{P_0}\right)} = \int_{T_0}^{T_C} \! \frac{\gamma}{T\left(\gamma-1\right)} \, \mathrm{d} T \,,
\label{eq:tc}
\end{align}
where $P_C$ is the measured EOC pressure, $T_C$ is the temperature at that time, $P_0$ and $T_0$ are the initial pressure and temperature, respectively, and $\gamma$ is the temperature-dependent ratio of specific heats of the mixture.

In the following, we assume that there is negligible change of the reactant mole fractions during compression and thus the specific heat ratio is a function of temperature only.
In addition, most mixtures studied in RCMs are composed primarily of \ce{O2}, \ce{N2}, and \ce{Ar} and typically no more than \SI{10}{\percent} of any mixture is the hydrocarbon fuel.
Since the specific heats of \ce{O2}, \ce{N2} and \ce{Ar} are only weakly temperature dependent over the range of temperatures experienced during compression, for the purposes of this uncertainty analysis, the total specific heat is approximated as a linear function of temperature:
\begin{align}
\hat{C}_{p} &\approx a + b T \,, \label{eq:cp-fit}
\end{align}
where $\hat{C}_{p}$ is the specific heat at constant pressure normalized by the universal gas constant, and $a$ and $b$ are found by fitting the total non-dimensional specific heat over the temperature range from \SIrange{300}{1100}{\kelvin}, as discussed in \cref{sec:unc-cp}.

With this approximation of the specific heat, \cref{eq:tc} can be analytically integrated to find the temperature at the end of compression:
\begin{align}
\label{eq:explicit-tc}
T_C = \frac{a \lam{\left(\frac{b}{a} \exp\!{\left[\frac{b T_0}{a}\right]} T_0 \left[\frac{P_C}{P_0}\right]^{\frac{1}{a}}\right)}}{b} \,,
\end{align}
where $\lam{(\ldots)}$ is Lambert's $\lam$ function \cite{Corless1996}.
Note that the following analysis is conducted for the temperature at the end of compression ($T_C$) because this is the reference temperature typically used to report ignition delays.
Nevertheless, the analysis can be applied to any definition of a reference temperature with suitable modifications.

After the end of compression, the piston is locked in place and the reactions proceed in the constant volume chamber.
During this induction period, the reactants are losing energy to the relatively cold reactor walls by heat transfer.
This heat transfer causes a decrease in the temperature and corresponding decrease in the pressure.
The temperature during this induction period can be computed under the adiabatic core hypothesis by modeling the heat loss as an adiabatic, reversible volume expansion.
We note that this procedure breaks down under conditions where the adiabatic core hypothesis no longer holds; that is, when significant changes in chemical enthalpy occur (e.g.\ ignition, including first-stage ignition, begins) or the boundary layer consumes the core.

In addition, the mole fractions of the species (particularly the fuel) may be changing during this time, such that the assumption of constant species mole fractions made previously in this section no longer applies.
Therefore, the specific heat may no longer be a function of temperature alone, but instead a function of temperature and composition.
This additional dependence substantially complicates the analysis of the temperature and its uncertainty during the post-compression period, but can be accounted for by using simulations with detailed chemistry included.
This procedure has been validated experimentally by \citet{Das2012a}, who showed that the temperature of the mixture measured by two-line infrared thermometry closely matched with the temperature computed by the adiabatic core hypothesis (including detailed chemistry) during the induction period.

\section{Uncertainty of Compressed Temperature}
\label{sec:unc-TC}

In this work, two uncertainty analysis methods are used to estimate the uncertainty in the compressed temperature.
The first method is an estimation of the uncertainty assuming that each parameter is independent.
In the method of independent parameters, the uncertainty of $a$ and $b$ are considered explicitly.
There are thus five parameters in this method---$T_0$, $P_0$, $P_C$, $a$, and $b$---three of which are the same as in the Monte Carlo method.
This method is called the method of independent parameters and is described in \cref{sec:independent-parameters}.

The second is a Monte Carlo method that directly uses \cref{eq:explicit-tc} and considers three major parameters---$T_0$, $P_0$, and $P_C$---whose uncertainties must be estimated.
In addition, $a$ and $b$ must be calculated while taking into account the uncertainty of the composition of the mixture; that is, the uncertainty of the specific heat must be accounted for in its fitting parameters in the Monte Carlo method.
However, there is no explicit uncertainty associated with $a$ and $b$ per se in this method.
The Monte Carlo analysis is described in \cref{sec:monte-carlo}.

Two Python scripts have been developed to perform the each analysis, one each for liquid fuels with the injected mass filling method and one each for gaseous fuels (or volatile liquid fuels with the partial pressure filling method), for a total of four scripts.
The scripts are available in the supplemental material and are also available on Zenodo at \url{http://dx.doi.org/10.5281/zenodo.11968}.
The scripts use the NumPy \cite{Oliphant2007} and SciPy \cite{scipy} packages for Python along with the Cantera software \cite{cantera}.
In this work, versions 3.4.1 of Python, 1.8.2 of NumPy, 0.14.0 of SciPy, and 2.1.1 of Cantera are used.

\subsection{Independent Parameters Methodology}
\label{sec:independent-parameters}

In the method of independent parameters, the uncertainty in $T_C$ is estimated by the quadratic sum of the standard deviation in the parameters in \cref{eq:explicit-tc} multiplied by the partial derivative of \cref{eq:explicit-tc} with respect to each of the parameters \cite{Taylor1982}:
\begin{equation}
\label{eq:unc-taylor}
\delta_{T_C} = 2\cdot\sqrt{\sum_{\substack{k=T_0,\ P_0,\\ P_C,\ a,\ b}} \left(\frac{\partial T_C}{\partial k}\right)^2 \sigma_{k}^2} \,,
\end{equation}
where $\delta_{T_C}$ is the total uncertainty in $T_C$, $k$ is a parameter such that the sum is over the five parameters in the independent parameters analysis, and $\sigma_k$ is the standard deviation of the parameter $k$.
The coverage factor of $\delta_{T_C}$ is $2$ such that the confidence interval is \SI{95.4}{\percent} if the distribution of the total uncertainty is assumed to be normal.

The uncertainties of three of the parameters ($\sigma_{T_0}$, $\sigma_{P_0}$, and $\sigma_{P_C}$; see \cref{sec:unc-parameters}) are estimated from manufacturer's specifications.
Note that in the method of independent parameters the uncertainties of $a$ and $b$ ($\sigma_a$ and $\sigma_b$) are considered directly, unlike in the Monte Carlo analysis.
Therefore, the uncertainties of $a$ and $b$ are estimated following the procedure of \citet{York2004}, detailed in the Supplementary Material, Appendix B.

\subsection{Monte Carlo Methodology}
\label{sec:monte-carlo}

For a liquid fuel, the injected (nominal) mass of fuel, mixing tank volume, ambient temperature, and the relative proportions of \ce{O2}, \ce{N2}, and \ce{Ar} are input to the script.
The proportions of the gases and the nominal mass of fuel are used to compute the nominal partial pressure of each gas in the mixing tank.
For a gaseous fuel, the nominal pressures after filling each gas are directly input to the script and the nominal partial pressure is calculated from this information.
Normal distributions are created for each of the gases in the order that they were filled into the mixing tank.
For instance, if the order of filling is \ce{O2}, \ce{N2}, then \ce{Ar} the following procedure is used:
\begin{enumerate}
\item The center of the normal distribution for \ce{O2} is set at the nominal partial pressure for the \ce{O2}, with standard deviation as computed in \cref{sec:unc-gascomp}.
\item The pressure of \ce{O2} for a given simulated run is found by sampling the inverse of the cumulative distribution function, known as the percent point function (PPF), of the \ce{O2} distribution at a random location between \num{0} and \num{1}.
\item The center of the normal distribution for \ce{N2} is set at the calculated nominal partial pressure for the \ce{N2} plus the sampled value of the pressure of \ce{O2}. In this way, the uncertainty in the setting for \ce{O2} is coupled to the uncertainty for \ce{N2}. The standard deviation of the distribution is as computed in \cref{sec:unc-gascomp}.
\item The pressure of \ce{N2} for a given simulated run is found by sampling the PPF of the \ce{N2} distribution at a random location between \num{0} and \num{1}.
\item The center of the normal distribution for \ce{Ar} is set at the calculated nominal partial pressure for the \ce{Ar} plus the sampled value of the pressure of \ce{N2}+\ce{O2}, with standard deviation as computed in \cref{sec:unc-gascomp}.
\item The pressure of \ce{Ar} for a given simulated run is found by sampling the PPF of the \ce{Ar} distribution at a random location between \num{0} and \num{1}.
\end{enumerate}

For gaseous fuels, the pressures sampled from the distributions are converted into sampled partial pressures.
These partial pressures are used to compute the mole fraction of each component by Dalton's Law \cite{Dalton1801,Gillespie1930}.
For gaseous fuels, the mole fractions are thus independent of the volume of the mixing tank and the ambient temperature measurement.
For liquid fuels considered under the injected mass filling procedure, the sampled partial pressure of each gaseous component is converted to a number of moles in the mixing tank by the ideal gas law.
The ambient temperature and volume of the mixing tank used in the ideal gas law are sampled from the PPF of distributions of the temperature and mixing tank volume respectively.
The distribution of the temperature is as described in \cref{sec:unc-T0}; the distribution of the mixing tank volume is as described in \cref{sec:unc-V0}.
The number of moles of fuel is calculated from the mass of fuel sampled from the PPF for the mass.

The mole fraction of each component is input to an instance of the Cantera \verb|Solution| class.
The Cantera \verb|Solution| is used to calculate the non-dimensional specific heat as a function of temperature.
A line is fit to the specific heat using the NumPy function \verb|polyfit|.
For the mixtures considered in this method, the correlation coefficient for the linear fits are greater than \num{0.95}, indicating a good fit.
Finally, the initial pressure, initial temperature of the compression, and compressed pressure are input to the script and \cref{eq:explicit-tc} is used to calculate $T_C$.
This procedure is able to capture the correlation between the uncertainty of the gases as they are filled by simulating the filling process.
Moreover, using this procedure, it is possible to simulate several uncertainty distributions for any of the parameters and determine the effect of the uncertainty distribution on the total uncertainty.

The sampling and computation of $T_C$ are run $\num{E6}$ times and the mean and standard deviation are extracted from the calculated values of $T_C$. The reported uncertainty (called $s$) is twice the standard deviation from this analysis, thus representing an \SI{95.4}{\percent} confidence interval for normally distributed uncertainty.

\section{Uncertainty Distributions of the Parameters}
\label{sec:unc-parameters}

A general procedure to determine the uncertainties of each of the parameters and the total specific heat are described in the following.
For specific values used in the analysis in \cref{sec:results} for the RCM at UConn, the reader is referred to the Supplementary Material, Appendix A.
Similar analysis can be conducted for other machines.
In the following, the manufacturer's quoted uncertainty is assumed to represent a normal distribution with a coverage factor of two, unless otherwise specified.
The uncertainty is divided by the coverage factor to determine the standard deviation, as recommended by the British Measurement and Testing Association Measurement Good Practice Guide No.\ 36 \cite{Birch2001}.

\subsection{Uncertainty of the Measured Pressures}
\label{sec:unc-p0}\label{sec:unc-PC}

A static transducer is generally used to measure the initial pressure prior to each experiment and also the pressure of the reactants as they are filled into the mixing tank (see \cref{sec:unc-cp}).
In addition, a dynamic piezoelectric transducer is typically used to measure the pressure during and after the compression stroke.
The charge signal from the dynamic transducer is converted to a voltage signal by a charge amplifier, which is in turn measured by a computerized data acquisition system.
The uncertainty of the pressure transducers are typically specified as a percentage of the full scale range.
The precision uncertainty of the display used to read the pressure and the uncertainty in the signal acquisition equipment may be negligible compared to this uncertainty.
In this case, the uncertainty of the pressures are assumed to be normally distributed with mean at the measured pressure and standard deviation of one half of the manufacturers specification.
Otherwise, the uncertainty of the display and any signal acquisition equipment needs to be considered by the sum in quadrature of all of the uncertainties.
The standard deviation of the initial pressure measurement is denoted by $\sigma_{P_0}$ and the standard deviation of the compressed pressure measurement is denoted by $\sigma_{P_C}$ in the following.



\subsection{Uncertainties of the Initial and Ambient Temperatures}
\label{sec:unc-T0}

Two temperatures are measured in the course of RCM experiments.
The first is the temperature at which the gases are filled in to the mixing tank---known as the ambient temperature ($T_a$)---and the second is the initial temperature of the reaction chamber prior to compression---known as the initial temperature ($T_0$).
The uncertainty in the measured temperature is due to the standard limits of error of the thermocouples used to measure the temperature.
The limits of error of the standard thermocouple types are prescribed by the ASTM standard E230 \cite{ASTM2012}.
Additional sources of uncertainty include the cold junction compensation, A/D converter, and the display on the process meter used to read the temperature.
These uncertainties may be negligible compared to the uncertainty due to the thermocouple specification.

There is some ambiguity about how to interpret the uncertainty defined in the ASTM standard.
Thus, three uncertainty distributions are studied in this work for the measured temperatures.
The objective is to demonstrate the effect of varying distributions due to the sensitivity of the overall uncertainty to the uncertainty in the thermocouple measurements (see \cref{sec:results}).
The first distribution is a normal distribution assuming the tolerance is given with a coverage factor of two; the uncertainty in this case for the independent parameters analysis is $\sigma_T = \mathrm{LOE}/2$, where $\mathrm{LOE}$ is the limit of error as specified in the ASTM standard.
The second distribution is a triangular distribution with the limits at the limits of error according to the ASTM standard; the uncertainty in this case for the independent parameters analysis is $\sigma_T = \mathrm{LOE}/\sqrt{6}$  as recommended by the Guide to Uncertainty in Measurements (GUM) \cite{JCGM2008}.
The final distribution is a uniform distribution with the limits at the limits of error according to the ASTM standard; the uncertainty in this case for the independent parameters analysis is $\sigma_T = \mathrm{LOE}/\sqrt{3}$ as recommended by the GUM \cite{JCGM2008}.
In general, the uncertainty in the initial and ambient temperatures will be equal ($\sigma_{T} = \sigma_{T_0} = \sigma_{T_a}$) but it is possible to vary them independently, e.g.\ by using different thermocouples for the measurements.

\subsection{Uncertainty of Liquid Fuel Mass}
\label{sec:unc-liquid}

According to the procedure defined in \cref{sec:test-gas} for liquid fuels, the mass of the syringe used for injection is measured twice---once before and once after fuel injection---with the difference in mass taken as the amount of fuel injected.
The uncertainty in the measured mass has three components---the precision of the scale, the deviation from linearity, and the repeatability of the measurement.
These uncertainties are assumed to be normally distributed with coverage factors of 2 such that the standard deviations are one half the specified values.
Thus, the total standard deviation in the mass of the liquid fuel may be estimated by the sum in quadrature of these values \cite{Taylor1982}.
The standard deviation of the fuel mass measurement is represented by $\sigma_m$ in the following.

\subsection{Uncertainty of Gaseous Components}
\label{sec:unc-gascomp}

All filling of gases is done at room temperature as measured by one of the thermocouples described in \cref{sec:unc-T0}.
If a liquid fuel is used, the nominal pressure of each of the gaseous components in the oxidizer is computed by application of the ideal gas law after the injected mass of liquid fuel is determined.
Otherwise, if a gaseous fuel is used, the nominal pressure of each component is computed prior to the start of filling.
Since the same pressure transducer is used to measure the gas pressure during filling of the mixing tank as during measurement of the initial pressure, the same standard deviation is used, $\sigma_{P_i} = \sigma_{P_0}$.
In the Monte Carlo analysis, the mean of the distribution is set at the value randomly sampled from the cumulative distribution function for the previous component added to the nominal value of the required pressure (see \cref{sec:monte-carlo}).
Because of this correlation in the uncertainties, $\sigma_{P_i}$ is the standard deviation of the measurement of the total pressure, not necessarily the standard deviation of the measurement of the partial pressure of component $i$ in the Monte Carlo analysis.
Since the width of the distribution is constant (set by the standard deviation $\sigma_{P_i}$) while the mean of the distribution is increasing for each gas, the ratio of the standard deviation to the mean of the distribution is smaller for gases that are filled in later in the filling process.
This correlation between the pressures of the gaseous components is not captured by the independent parameters analysis; if the effect of the correlation is large, an analysis such as the Monte Carlo analysis conducted in this study is required.

\subsection{Uncertainty of the Mixing Tank Volume}
\label{sec:unc-V0}

The volumes of the mixing tanks described in \cref{sec:test-gas} are measured by filling them with water at a known temperature.
The mass of the water is measured and the density is used to convert the mass to the volume.
The uncertainty in the mass of water filled in to the tank can be estimated by considering the sum in quadrature of the uncertainties associated with the scale used to measure the mass, similar to the liquid fuel measurements (see \cref{sec:unc-liquid}).
The volume is computed by dividing the measured mass by the density of the water, $V = m_{\text{water}}/\rho_{\text{water}}$.
The density of the water can be estimated from a number of sources, including the NIST Chemistry WebBook \cite{Lemmon2014,Wagner2002}, and must take into account any uncertainty in the  ambient temperature measurement.
The distribution of the uncertainty is assumed to be normal with mean at the computed value of the volume and standard deviation of $\sigma_V$.

\subsection{Uncertainty of the Total Specific Heat}
\label{sec:unc-cp}

Since each reactant is filled in to the mixing tank sequentially, the uncertainties of the amounts of each of the reactants are correlated.
As discussed previously in \cref{sec:test-gas}, the order of filling is liquid fuel (if any), oxidizer, and gaseous fuel (if any).
The order of filling and the total pressure in the mixing tank after each gaseous component is filled are recorded.

Two fuels are studied in this analysis, methylcyclohexane (MCH) and prop\-ene.
These represent, respectively, a liquid fuel and a gaseous fuel.
The thermodynamic polynomials for the species under consideration in this work are taken from the work of \citet{Weber2014} for MCH, \ce{O2}, \ce{N2}, and \ce{Ar}, and the work of \citet{Burke2015} for propene.
In this work, it is assumed that the uncertainty in the specific heat of each species is negligible.
This is considered an acceptable assumption for stable species such as fuel molecules, oxygen, nitrogen, argon, and other common diluents.
Experience with several thermodynamic databases has shown that the typical variation in the full polynomial fit to $\hat{C}_p$ between databases causes approximately \SI{1}{\kelvin} changes in $T_C$.

Extending this analysis to fuel blends with several components is straightforward.
The user need only specify the uncertainty of the fraction of the fuel components in the blend, in addition to the uncertainties discussed in this section.
It should be noted that the uncertainty in the fractions of the components in a multi-component fuel blend is inherently correlated and the Monte Carlo method should be preferred for these analyses.
In this work, we consider only the single-component fuels discussed previously.

In this uncertainty analysis, the specific heat is fit by a linear relationship as discussed in \cref{sec:unc-TC}.
The end points of this fit are chosen to be \SIlist{300;1100}{\kelvin}.
The lower temperature is chosen as the lowest temperature over which the NASA polynomials are specified in the thermodynamic databases used for this study; the upper temperature is chosen to be above the highest compressed temperature expected in the analysis.

\section{Results and Discussion}
\label{sec:results}

\subsection{Conditions of Analysis}

\subsubsection{Methylcyclohexane}

The following analyses are conducted for mixtures of MCH with \ce{O2}/\ce{N2}/\ce{Ar}.
Three mixtures are studied whose proportions are given in \cref{tab:mch-mixes}.
The proportions of the reactants match the conditions studied in \citet{Weber2014}.
These mixtures are studied for two sets of initial/compressed conditions each, as shown in \cref{tab:mch-cases}.
All six cases are based on experimental conditions from the work of \citet{Weber2014}, who used the larger mixing tank for all the mixtures ($V = \SI[separate-uncertainty]{16.60(1)}{\liter}$).

\subsubsection{Propene}

The following analysis is conducted for several mixtures of propene (\ce{C3H6}) in \ce{O2}/\ce{N2}/\ce{Ar}.
The mixtures studied here match some of those studied in the work of \citet{Burke2015}.
The nominal partial pressures of each component in the mixtures studied in this section are given in \cref{tab:propene-mixes}.
Mix 1 has \ce{O2}/diluent ratio of $\num{1}:\num{3.76}$, mixes \numlist{2;3} are made with \SI{4}{\percent} \ce{O2} in the mixture, and mixes \numrange{4}{6} are made with \SI{12}{\percent} \ce{O2} in the mixture.
The initial and compressed conditions of the simulations are shown in \cref{tab:propene-cases}.
The compressed conditions include relatively low pressures near \SI{10}{\bar} and higher pressures near \SI{40}{\bar}.
All eight cases are based on experimental conditions from the work of \citet{Burke2015}.

\begin{table}
\centering
\caption{Molar proportions of the MCH mixtures studied in this work, matching the conditions in the work of \citet{Weber2014}.}
\label{tab:mch-mixes}
\begin{tabular}{*{6}{S}}
    \toprule
    {Mix \#} & {$\phi$} & {MCH} & {\ce{O2}} & {\ce{N2}} & {\ce{Ar}} \\
    \midrule
    1 & 1.0 & 1 & 10.5 & 12.25 & 71.75 \\
    2 & 0.5 & 1 & 21.0 &  0.00 & 73.50 \\
    3 & 1.5 & 1 &  7.0 & 16.35 & 71.15 \\
    \bottomrule
\end{tabular}
\end{table}

\pgfplotstableread[
    col sep=&,
]{mch-cases.dat}\mchcases

\begin{table}
\centering
\caption{Initial/compressed conditions for the estimation of the uncertainty of $T_C$ in MCH experiments.}
\label{tab:mch-cases}
\pgfplotstabletypeset[
    columns/{Case}/.style={
        string type,
    },
    columns/MixNo/.style={
        column name={Mix \#},
    },
    columns/P0/.style={
        column name={$P_0$ [\si{\pascal}]},
        sci, sci zerofill,
        precision=4,
    },
    columns/T0/.style={
        column name={$T_0$ [\si{\kelvin}]},
    },
    columns/PC/.style={
        column name={$P_C$ [\si{\pascal}]},
        sci zerofill,
    },
    columns/mfuel/.style={
        column name={$m_{\text{fuel}}$ [\si{\gram}]},
        fixed zerofill,
    },
    columns/Ta/.style={
        column name={$T_a$ [\si{\degreeCelsius}]},
        fixed zerofill,
        precision=1,
    },
    columns={Case, MixNo, P0, T0, PC, mfuel, Ta},
]\mchcases
\end{table}

\pgfplotstableread[
    col sep=comma,
]{propene-mixes.dat}\propenemixes

\begin{table}
\centering
\caption{Nominal partial pressures of the components studied in the propene experiments.}
\label{tab:propene-mixes}
\pgfplotstabletypeset[
    columns/MixNo/.style={
        column name={Mix \#},
    },
    columns/phi/.style={
        column name={$\phi$},
        fixed zerofill,
        precision=1,
    },
    columns/o2/.style={
        column name={\ce{O2} [\si{\torr}]},
    },
    columns/n2/.style={
        column name={\ce{N2} [\si{\torr}]},
    },
    columns/ar/.style={
        column name={\ce{Ar} [\si{\torr}]},
    },
    columns/fuel/.style={
        column name={\ce{C3H6} [\si{\torr}]},
    },
]\propenemixes
\end{table}

\pgfplotstableread[
    col sep=&,
]{propene-cases.dat}\propenecases

\begin{table}
\centering
\caption{Initial/compressed conditions for the estimation of the uncertainty of $T_C$ in propene experiments.}
\label{tab:propene-cases}
\pgfplotstabletypeset[
    columns/{Case}/.style={
        string type,
    },
    columns/MixNo/.style={
        column name={Mix \#},
    },
    columns/P0/.style={
        column name={$P_0$ [\si{\pascal}]},
        sci, sci zerofill,
        precision=4,
    },
    columns/T0/.style={
        column name={$T_0$ [\si{\kelvin}]},
    },
    columns/PC/.style={
        column name={$P_C$ [\si{\pascal}]},
        sci zerofill,
    },
    columns={Case, MixNo, P0, T0, PC},
]\propenecases
\end{table}

\pgfplotstableread{mch-normal-results.dat}\mchnormalresults

\begin{table}
\centering
\begin{threeparttable}
\caption{Results of analysis for MCH using normal distributions for all the parameters.}
\label{tab:mch-results}
\pgfplotstabletypeset[
    every head row/.append style={
        after row/.add={%
            \arraybackslash%
            & $T_C$ [\si{\kelvin}]%
            & $T_C$ [\si{\kelvin}]%
            & [\si{\kelvin}]%
            &%
            & $T_C$ [\si{\kelvin}]%
            & [\si{\kelvin}]%
            &\\%
            }{},
    },
    columns/{Case}/.style={
        string type,
    },
    columns/{Reported TC}/.style={
        column name={Reported},
        fixed zerofill,
    },
    columns/{Mean Normal Temperature}/.style={
        precision=2,
        column name={Mean Calc.},
        fixed zerofill,
    },
    columns/{2 sigma norm}/.style={
        precision=2,
        column name={$s$},
        fixed zerofill,
    },
    columns/{Ind Param TC}/.style={
        precision=2,
        column name={Indep.\ Param.},
        fixed zerofill,
    },
    columns/{delta TC normal}/.style={
        precision=2,
        column name={$\delta_{T_C}$},
        fixed zerofill,
    },
    columns/{sigma percent norm}/.style={
        precision=2,
        column name={\si{\percent}\tnote{2}},
        fixed, fixed zerofill,
    },
    columns/{ind percent norm}/.style={
        precision=2,
        column name={\si{\percent}\tnote{1}},
        fixed, fixed zerofill,
    },
    columns={{Case}, {Reported TC}, {Ind Param TC}, {delta TC normal}, {ind percent norm}, {Mean Normal Temperature}, {2 sigma norm}, {sigma percent norm}},
]\mchnormalresults
\begin{tablenotes}
\item[1] $\si{\percent} = \delta_{T_C}/\text{Indep.\ Param.\ }T_C \times 100\si{\percent}$
\item[2] $\si{\percent} = s/\text{Mean Calc.\ }T_C \times 100\si{\percent}$
\end{tablenotes}
\end{threeparttable}
\end{table}

\pgfplotstableread{propene-normal-results.dat}\propenenormalresults

\begin{table}
\centering
\begin{threeparttable}
\caption{Results of analysis for propene using normal distributions for all the parameters.}
\label{tab:propene-results}
\pgfplotstabletypeset[
    every head row/.append style={
        after row/.add={%
            \arraybackslash%
            & $T_C$ [\si{\kelvin}]%
            & $T_C$ [\si{\kelvin}]%
            & [\si{\kelvin}]%
            &%
            & $T_C$ [\si{\kelvin}]%
            & [\si{\kelvin}]%
            &\\%
            }{},
    },
    columns/{Case}/.style={
        string type,
    },
    columns/{Reported TC}/.style={
        column name={Reported},
        fixed zerofill,
    },
    columns/{Mean Normal Temperature}/.style={
        precision=2,
        column name={Mean Calc.},
        fixed zerofill,
    },
    columns/{2 sigma norm}/.style={
        precision=2,
        column name={$s$},
        fixed zerofill,
    },
    columns/{Ind Param TC}/.style={
        precision=2,
        column name={Indep.\ Param.},
        fixed zerofill,
    },
    columns/{delta TC normal}/.style={
        precision=2,
        column name={$\delta_{T_C}$},
        fixed zerofill,
    },
    columns/{sigma percent norm}/.style={
        precision=2,
        column name={\si{\percent}\tnote{2}},
        fixed, fixed zerofill,
    },
    columns/{ind percent norm}/.style={
        precision=2,
        column name={\si{\percent}\tnote{1}},
        fixed, fixed zerofill,
    },
    columns={{Case}, {Reported TC}, {Ind Param TC}, {delta TC normal}, {ind percent norm}, {Mean Normal Temperature}, {2 sigma norm}, {sigma percent norm}},
]\propenenormalresults
\begin{tablenotes}
\item[1] $\si{\percent} = \delta_{T_C}/\text{Indep.\ Param.\ }T_C \times 100\si{\percent}$
\item[2] $\si{\percent} = s/\text{Mean Calc.\ }T_C \times 100\si{\percent}$
\end{tablenotes}
\end{threeparttable}
\end{table}

\pgfplotstableread{mch-distribution-results.dat}\mchdistresults

\begin{table}
\centering
\begin{threeparttable}
\caption{Results for varying probability distributions of thermocouples for case d in the MCH analysis.}
\label{tab:mch-thermocouples}
\pgfplotstabletypeset[
    columns={Distribution, {Independent Parameters}, {delta dist percent}, {Monte Carlo}, {sigma dist percent}},
    columns/Distribution/.style={
        string type,
    },
    columns/{Monte Carlo}/.style={
        precision=2,
        column name={$s$ [\si{\kelvin}]},
        fixed zerofill,
    },
    columns/{sigma dist percent}/.style={
        precision=2,
        column name={\si{\percent}\tnote{2}},
        fixed zerofill,
    },
    columns/{Independent Parameters}/.style={
        precision=2,
        column name={$\delta_{T_C}$ [\si{\kelvin}]},
        fixed zerofill,
    },
    columns/{delta dist percent}/.style={
        precision=2,
        column name={\si{\percent}\tnote{1}},
        fixed zerofill,
    },
]\mchdistresults
\begin{tablenotes}
\item[1] $\si{\percent} = \delta_{T_C}/\text{Indep.\ Param.\ }T_C \times 100\si{\percent}$
\item[2] $\si{\percent} = s/\text{Mean Calc.\ }T_C \times 100\si{\percent}$
\end{tablenotes}
\end{threeparttable}
\end{table}

\pgfplotstableread{propene-distribution-results.dat}\propenedistresults

\begin{table}
\centering
\begin{threeparttable}
\caption{Results for varying probability distributions of thermocouples for case e in the propene analysis.}
\label{tab:propene-thermocouples}
\pgfplotstabletypeset[
    columns={Distribution, {Independent Parameters}, {delta dist percent}, {Monte Carlo}, {sigma dist percent}},
    columns/Distribution/.style={
        string type,
    },
    columns/{Monte Carlo}/.style={
        precision=2,
        column name={$s$ [\si{\kelvin}]},
        fixed zerofill,
    },
    columns/{sigma dist percent}/.style={
        precision=2,
        column name={\si{\percent}\tnote{2}},
        fixed zerofill,
    },
    columns/{Independent Parameters}/.style={
        precision=2,
        column name={$\delta_{T_C}$ [\si{\kelvin}]},
        fixed zerofill,
    },
    columns/{delta dist percent}/.style={
        precision=2,
        column name={\si{\percent}\tnote{1}},
        fixed zerofill,
    },
]\propenedistresults
\begin{tablenotes}
\item[1] $\si{\percent} = \delta_{T_C}/\text{Indep.\ Param.\ }T_C \times 100\si{\percent}$
\item[2] $\si{\percent} = s/\text{Mean Calc.\ }T_C \times 100\si{\percent}$
\end{tablenotes}
\end{threeparttable}
\end{table}

\subsection{Independent Parameters Analysis}
\label{sec:mch-indep}

The results from the independent parameters analysis assuming a normal distribution for the temperatures for MCH are shown in \cref{tab:mch-results} and those for propene are shown in \cref{tab:propene-results}.
The Reported $T_C$ is the value that was reported for the conditions associated with each case in \citet{Weber2014} for MCH and \citet{Burke2015} for propene, calculated by \cref{eq:tc}.
The Indep.\ Param.\ $T_C$ is the compressed temperature calculated by \cref{eq:explicit-tc} for the conditions in that case.
The Indep.\ Param.\ $T_C$ deviates slightly from the Reported $T_C$ due to the combined effects of the linear fit to the specific heat and the uncertainty in the mole fraction of the components; for the independent parameters analysis, the nominal value of the parameters are used to compute $T_C$ by \cref{eq:explicit-tc} except for $a$ and $b$, which are affected by the uncertainty of the components.
Nonetheless, the difference is quite minor, indicating the small effect of this assumption.
The $\delta_{T_C}$ column is the uncertainty in the temperature from \cref{eq:unc-taylor} and the first \si{\percent} column is the ratio of the uncertainty ($\delta_{T_C}$) to the Indep. Param. $T_C$ times \SI{100}{\percent}.

The magnitude of the uncertainty is quite similar between propene and MCH.
However, since the $T_C$ values are higher for propene, the uncertainty relative to the compressed temperature is smaller for propene than MCH.

Comparing cases a and b, c and d, and e and f for MCH, and a and b, c, d, e, and f, and g and h for propene shows the influence of the initial pressure ($P_0$) on the uncertainty.
Lower magnitudes of the initial pressure cause slightly higher magnitude of the uncertainty.
This result was noted by \citet{Weber2011}, who stated that ``the uncertainty is dependent on the actual value of the initial pressure... [t]ypically, higher initial pressures result in lower uncertainties in the compressed temperature.''
Nevertheless, the uncertainty relative to the compressed temperature is nearly identical between the cases.

The variation of the uncertainty in $T_C$ due to varying distributions for the measured temperatures is shown for MCH in \cref{tab:mch-thermocouples} for case d of \cref{tab:mch-cases,tab:mch-results} and for propene in \cref{tab:propene-thermocouples} for case e of \cref{tab:propene-results,tab:propene-cases}.
The triangular distribution has the lowest $\sigma_{T}$ (see \cref{sec:unc-T0}) and the lowest uncertainty; the uniform distribution has the highest $\sigma_{T}$ and the highest uncertainty for both fuels.

In addition, a sensitivity analysis of the uncertainty to the parameters for MCH and propene was conducted by varying the uncertainty of each parameter independently.
The uncertainty of each parameter was reduced by \SI{50}{\percent} compared to the nominal value.
This analysis shows that the uncertainty in the initial pressure ($\sigma_{P_0}$) has only a small influence on the total uncertainty, despite the influence of the magnitude of the initial pressure on the uncertainty.
Moreover, the uncertainty is not strongly sensitive to the uncertainty of any of the other parameters, except the initial temperature measurement ($\sigma_{T_0}$).
Reducing the uncertainty of the thermocouple used to measure the initial temperature $T_0$ by a factor of $2$ simulates replacing this thermocouple with one using the special tolerances for K-type thermocouples specified in the ASTM standard \cite{ASTM2012}.
By switching to the special tolerances for K-type thermocouples ($\sigma_{T_0} = \SI{0.55}{\degreeCelsius}$), the uncertainty $\delta_{T_C}$ is reduced by approximately \SI{2}{\kelvin} for all the cases.

\subsection{Monte Carlo Analysis}
\label{sec:mch-monte-carlo}

\begin{figure}
\centering
\includegraphics[width=90mm]{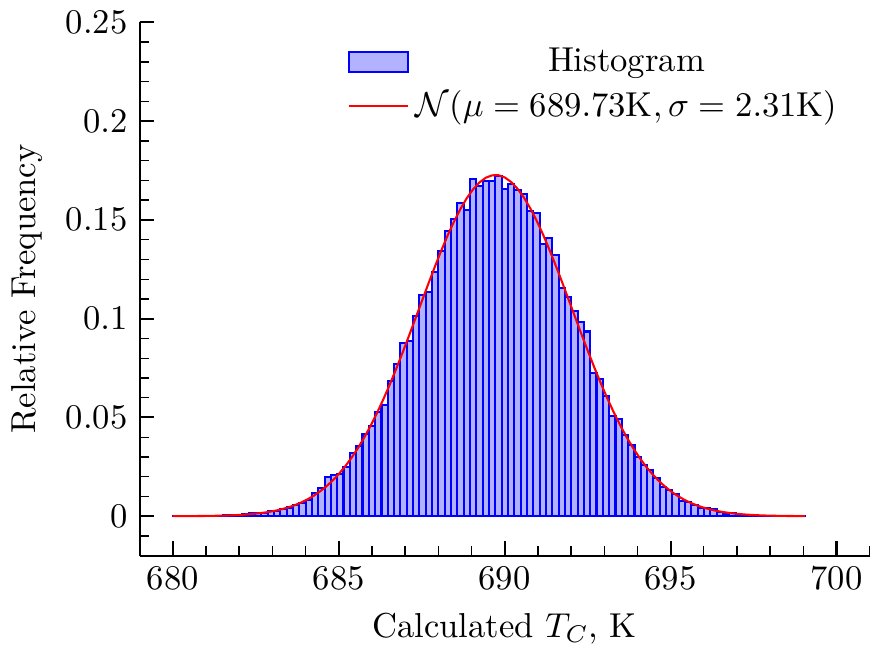}
\caption{Comparison of histogram of computed $T_C$ with a normal distribution ($\mathcal{N}$) with the same mean ($\mu$) and standard deviation ($\sigma=s/2$). The bin width of the histogram is approximately \SI{0.002}{\kelvin}. Conditions: MCH, case b.}
\label{fig:mch-histogram}
\end{figure}

\begin{figure}
\centering
\includegraphics[width=90mm]{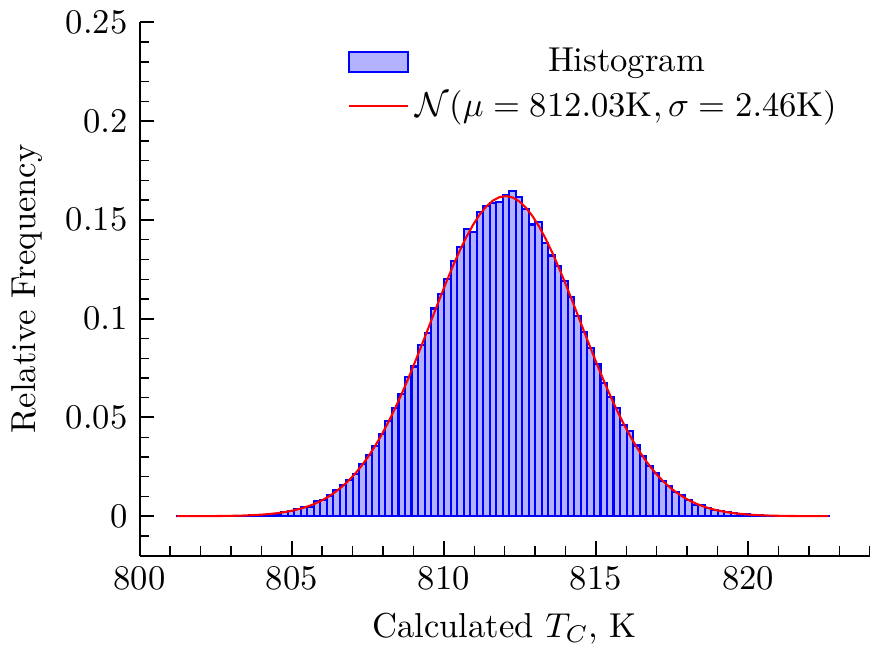}
\caption{Comparison of histogram of computed $T_C$ with a normal distribution ($\mathcal{N}$) with the same mean ($\mu$) and standard deviation ($\sigma=s/2$). The bin width of the histogram is approximately \SI{0.002}{\kelvin}. Conditions: propene, case f.}
\label{fig:propene-histogram}
\end{figure}

Using normal distributions for all of the parameters, the results of the Monte Carlo analysis for MCH are presented in \cref{tab:mch-results} and a histogram for a representative case is shown in \cref{fig:mch-histogram}.
In addition, the results of the Monte Carlo analysis assuming a normal distribution of the uncertainties of all the parameters are presented for propene in \cref{tab:propene-results} and a histogram for a representative case is shown in \cref{fig:propene-histogram}.
It can be seen from \cref{fig:mch-histogram,fig:propene-histogram} that a normal distribution with the same mean and standard deviation as the calculated $T_C$ closely follows the magnitude of the bins, indicating that twice the standard deviation is a good measure of the uncertainty of $T_C$.
References to the uncertainty of $T_C$ in the remainder of this section refer to twice the standard deviation, noted as $s$ in \cref{tab:mch-results,tab:mch-thermocouples,tab:propene-results,tab:propene-thermocouples}.
The mean temperature calculated from the Monte Carlo analysis is shown in \cref{tab:mch-results,tab:propene-results} in the Mean Calc.\ $T_C$ column.
The second \si{\percent} column is the ratio of $s$ to the Mean Calculated $T_C$ multiplied by \SI{100}{\percent}.
The Mean Calculated $T_C$ deviates slightly from the Reported $T_C$ due to the combined effects of the linear fit to the specific heat, the uncertainty in all of the parameters, and the correlation in the uncertainty of the mixture components.
Nonetheless, the difference is quite minor compared to the reported value, indicating the small effect of these assumptions on computing $T_C$.

Comparing the overall results for the Monte Carlo analysis and the independent parameters analysis shows that they are quite similar.
This indicates that the correlation of the uncertainties of the gaseous components has a relatively minor effect on the total uncertainty under the conditions of this analysis.
Comparing cases b, d, and f for MCH shows the influence of mixture composition on the standard deviation.
It is seen that changing the mixture composition has a relatively small effect on the standard deviation.
In addition, by comparing cases e and f for propene, the influence of the magnitude of the initial temperature can be estimated.
It is seen that the magnitude of the initial temperature has a small influence on the error.
Similar to the independent parameters analysis, the magnitude of the uncertainty is affected by the magnitude of the initial pressure ($P_0$), but the uncertainty relative to the Mean Calc.\ $T_C$ is not strongly affected by the magnitude of $P_0$.

Similar to the independent parameters case, a sensitivity analysis of the uncertainty $s$ to the uncertainties of the parameters was conducted by varying the uncertainty of each parameter independently.
Once again, the overall uncertainty $s$ is not sensitive to the standard deviation of any of the other parameters except the initial temperature.
Using the special tolerances for the initial temperature measurement ($\sigma_{T_0}=\SI{0.55}{\degreeCelsius}$ \cite{ASTM2012}) reduces the uncertainty by approximately \SI{2}{\kelvin} for all the cases.

\begin{figure}
\centering
\makebox[\textwidth][c]{\includegraphics[width=190mm]{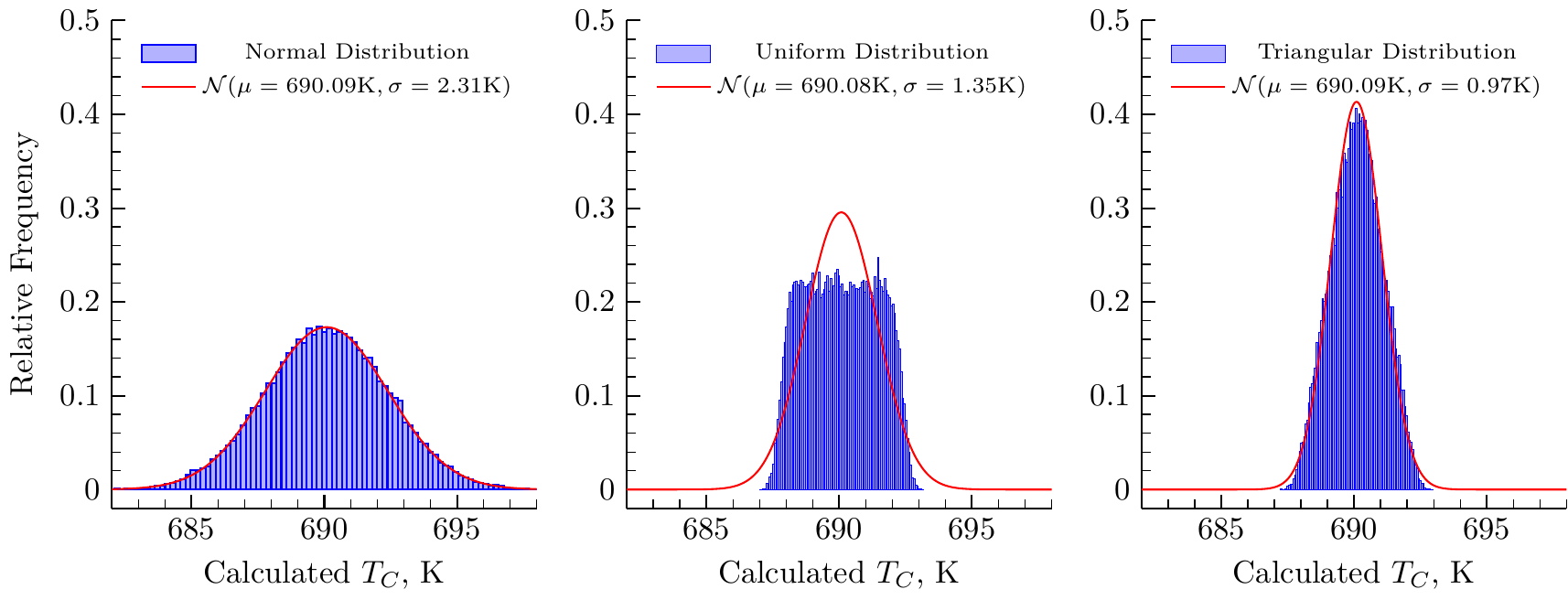}}
\caption{Comparison of the distribution of the calculated $T_C$ using three distributions for the uncertainty of the thermocouple measurement. The histograms match the conditions presented in \cref{tab:mch-thermocouples}. Normal, Uniform, and Triangular Distribution in the legend indicates that the respective distribution was used for the uncertainty of the thermocouple measurement. Conditions: MCH, case d}
\label{fig:mch-thermocouples}
\end{figure}

\begin{figure}
\centering
\makebox[\textwidth][c]{\includegraphics[width=190mm]{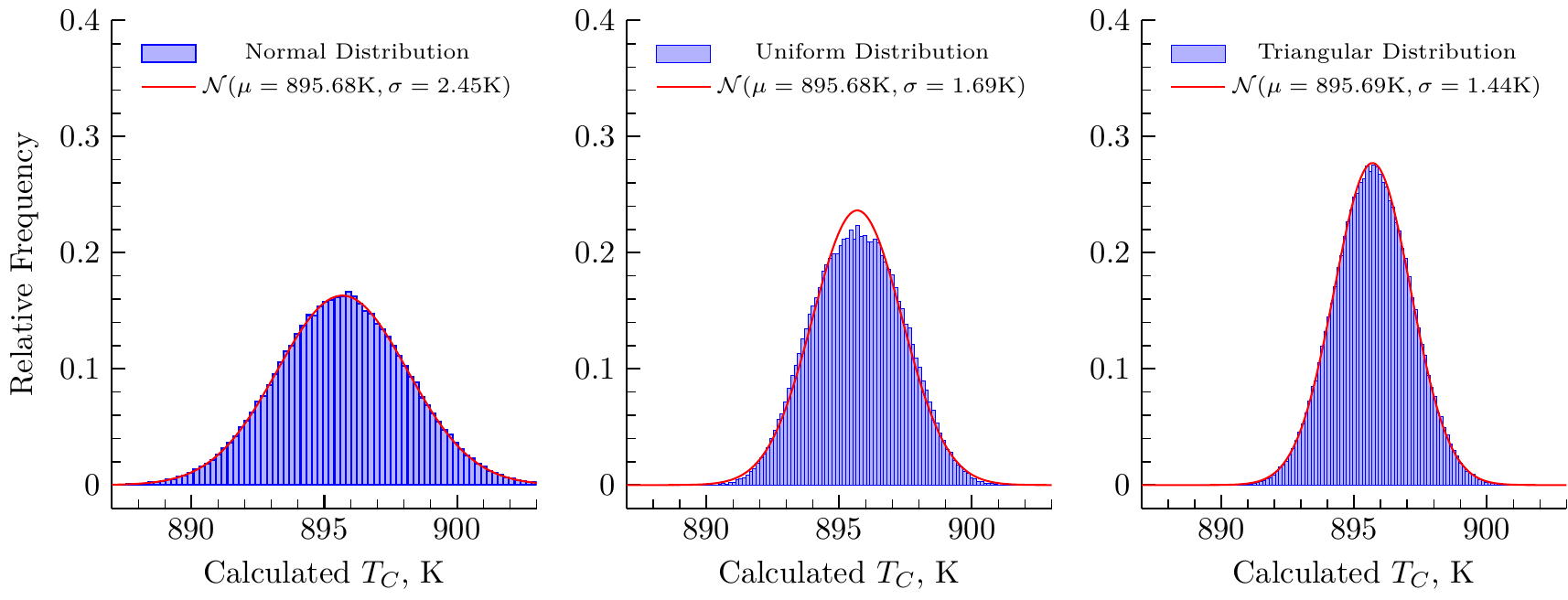}}
\caption{Comparison of the distribution of the calculated $T_C$ using three distributions for the uncertainty of the thermocouple measurement. The histograms match the conditions presented in \cref{tab:propene-thermocouples}. Normal, Uniform, and Triangular Distribution in the legend indicates that the respective distribution was used for the uncertainty of the thermocouple measurement. Conditions: propene, case e}
\label{fig:propene-thermocouples}
\end{figure}

\Cref{tab:mch-thermocouples,tab:propene-thermocouples} show a comparisons of the uncertainty of case d for MCH and case e for propene when assuming various distributions for the uncertainty of the thermocouples, and \cref{fig:mch-thermocouples,fig:propene-thermocouples} show a comparison of the histograms of the calculated $T_C$.
It can be seen in \cref{fig:mch-thermocouples,fig:propene-thermocouples} that, similar to \cref{fig:mch-histogram,fig:propene-histogram}, assuming a normal distribution for the uncertainty of the thermocouple measurements results in an normal distribution of $T_C$.
However, assuming a uniform distribution of the uncertainty of the thermocouple measurements strongly affects the resulting distribution of $T_C$ for MCH, but does not have as strong an influence on the total uncertainty of the propene case.
This may indicate the influence of the ambient temperature on the distribution of $T_C$ in the MCH case, since the ambient temperature is not a parameter in the propene case.
Note that the standard deviation as estimated in \cref{tab:mch-thermocouples,tab:propene-thermocouples} assumes a normal distribution of the calculated $T_C$; as such, the uncertainty in the ``Uniform'' case is slightly underestimated for MCH due to the non-normal distribution of $T_C$.
Nonetheless, the uncertainty in the ``Uniform'' case is less than the uncertainty in the ``Normal'' case for both MCH and propene, a result which differs from the independent parameters analysis.
Finally, assuming a triangular distribution of the uncertainty of the thermocouple measurements does not strongly affect the shape of the distribution of calculated $T_C$.
However, the standard deviation is reduced by a factor of approximately $2$ compared to the normal distribution for both fuels.

\section{Conclusions}
\label{sec:conclusions}

Rapid compression machines are an important platform for combustion experiments at high-pressure and low-to-intermediate-temperature conditions.
In addition to the primary data reported from RCMs (including ignition delays and intermediate species profiles), the temperature of the gas mixture in the reaction chamber is of considerable interest due to the exponential dependence of the reaction rates on the temperature.
The temperature of the gas mixture in the reaction chamber during and after compression is typically not measured.
Instead, the temperature is typically estimated by assuming an isentropic compression/expansion process.
The uncertainty associated with this procedure is affected by a number of systematic and random errors that must be investigated in detail to bound the uncertainty of the temperature estimation.
In this work, a methodology is developed to assess the uncertainty of the temperature estimation at the end of compression due to random errors in the experimental equipment.

Two methods have been developed for the analysis in this work, including a method that assumes all of the parameters are independent, and a Monte Carlo method to take into account the correlation of several of the pertinent parameters.
In addition, two fuels with different mixture preparation procedures are analyzed.
The first fuel, methylcyclohexane, is injected into the mixing tank in its liquid state, while the second fuel, propene, is added to the mixing tank as a gas.
These cases represent two of the possible mixture filling modes for the RCM and are the two used on the RCM at UConn for the studies of \citet{Weber2014} and \citet{Burke2015}.
The conditions at the end of compression considered for the uncertainty analysis of these fuels span a large portion of the operating regime of the RCM at UConn, including low temperatures near \SI{650}{\kelvin}, high temperatures near \SI{1050}{\kelvin}, low pressures near \SI{10}{\bar}, and high pressures near \SI{50}{\bar}.
Finally, three distributions are considered for the uncertainty of the measured initial and ambient temperatures.

Examining the results in detail shows some differences between the MCH cases and the propene cases.
In particular, varying the distribution of the uncertainty of the measured temperatures has different effects for propene and MCH in the Monte Carlo analysis.
In MCH, using uniform distributions of the uncertainties of the measured temperatures causes substantial changes in the shape of the distribution of the calculated $T_C$, but for propene, the same change has only a small effect on the shape of the distribution of the calculated $T_C$.
However, in the independent parameters analysis, varying the distribution of the measured temperatures show the same effect for both fuels.
In addition, the uncertainties of the MCH and propene cases from the Monte Carlo analysis and the independent parameters analysis are of similar magnitude.

There are other similarities between the MCH and propene results as well.
For instance, it is shown that the magnitude of the initial pressure ($P_0$) has a notable influence on the magnitude of the uncertainty for both fuels, with higher magnitudes of the initial pressure resulting in lower absolute uncertainty of $T_C$.
However, the uncertainty of the initial pressure measurement has only a small influence on the overall uncertainty.
Moreover, the uncertainties of the other parameters, except the temperature measurements, also have a small influence on the overall uncertainty.
By using the special tolerances of K-type thermocouples defined by ASTM, the uncertainty in the compressed temperature is reduced by approximately \SIrange{1}{2}{\kelvin}, which may be as much as \SI{20}{\percent} of the absolute uncertainty values in some cases.
Thus, one way to reduce the uncertainty in estimating $T_C$ is to improve the measurement of the initial and ambient temperatures.

The analyses in this study show that the relative uncertainty in the estimated temperature at the end of compression ($T_C$) does not exceed \SI{0.7}{\percent} of the mean calculated temperature for the instruments installed on the RCM at UConn.
These results provide a reference that can be used for future RCM studies at UConn.
Furthermore, the method and scripts developed in this work can be applied to other similar RCM facilities with only minor modifications.
Care must be taken to supply instrument specific uncertainties for each of the parameters and properly replicate the correlation of the parameters in the experimental procedure used on each machine.
Finally, the results show that the Monte Carlo and independent parameters analyses estimate similar uncertainties under the conditions of this work.
However, this may not be the case in general, so care should be exercised when using an analysis procedure that does not account for the correlation between the uncertainties of the parameters.

\section*{Acknowledgments}
This material is based upon work supported by the National Science Foundation under Grant No. CBET-1402231 and as part of the Combustion Energy Frontier Research Center, an Energy Frontier Research Center funded by the U.S. Department of Energy, Office of Science, Office of Basic Energy Sciences, under Award Number DE-SC0001198.

\appendix
\section{Supplementary Material}
Supplementary material associated with this article can be found in the online version.

\section*{References}
\bibliographystyle{elsarticle-num-CNF}
\bibliography{extracted}
\end{document}